\documentclass[a4paper,11pt]{article}
\usepackage{pos}
\usepackage{placeins}

\title{Determination of Lee-Yang edge singularities in QCD by rational approximations}

\author*[a]{K. Zambello}
\author[c]{D. A. Clarke}
\author[b]{P. Dimopoulos}
\author[b]{F. Di Renzo}
\author[d]{J. Goswami}
\author[c]{G. Nicotra}
\author[c]{C. Schmidt}
\author[b]{S. Singh}

\affiliation[a]{Dipartimento di Fisica, Universit\`a di Pisa and INFN, Sezione di Pisa, Pisa, Italy}
\affiliation[b]{Dipartimento di Scienze Matematiche, Fisiche e Informatiche, Universit\`a di Parma and INFN, Gruppo Collegato di Parma, Parma, Italy}
\affiliation[c]{Fakult\"at f\"ur Physik, Universit\"at Bielefeld, D-33615, Bielefeld, Germany}
\affiliation[d]{RIKEN Center for Computational Science, Kobe 650-0047, Japan}

\emailAdd{kevin.zambello@pi.infn.it}
\emailAdd{clarke.davida@gmail.com}
\emailAdd{petros.dimopoulos@unipr.it}
\emailAdd{francesco.direnzo@unipr.it}
\emailAdd{jishnu@physik.uni-bielefeld.de}
\emailAdd{gnicotra@physik.uni-bielefeld.de}
\emailAdd{schmidt@physik.uni-bielefeld.de}
\emailAdd{simran.singh@unipr.it}

\abstract{We report updated results on the determination of Lee-Yang edge (LYE) singularities in $N_f=2+1$ QCD using
highly improved staggered quarks (HISQ) with physical masses on $N_\tau = 4, 6, 8$ lattices.
The singularity structure of QCD in the complex $\mu_B$ plane is probed using conserved charges calculated at imaginary $\mu_B$.
The location of the singularities is determined by studying the (uncancelled) poles of multi-point Pad\'e approximants. 
We show that close to the Roberge-Weiss (RW) transition, the location of the LYE singularities scales according to the $3$-$d$ $Z(2)$
universality class. By combining the new $N_\tau = 6$ data with the $N_\tau = 4$ data from our previous analysis we extract a
rough estimate for the RW temperature in the continuum limit. We also discuss some preliminary results for the singularities
close to the chiral phase transition obtained from simulations on $N_\tau = 6,8$ lattices.}

\FullConference{%
The 39th International Symposium on Lattice Field Theory,\\
8th-13th August, 2022,\\
Rheinische Friedrich-Wilhelms-Universität Bonn, Bonn, Germany
}

\begin{document}
\maketitle

\section{Introduction}
A detailed knowledge of the phase diagram of QCD is necessary for a complete understanding of the physics of strongly
interacting matter at finite temperature and density. 
At zero chemical potentials the QCD transition is known to be a crossover. The transition is conjectured to
become a first-order transition at large chemical potentials; then one expects the presence of a second order critical endpoint
(CEP), whose exact location however is still unknown.
Much progress has been done in investigating the QCD phase diagram at small chemical potentials (for a review see 
ref. \cite{Guenther:2022wcr}). Unfortunately at large chemical potentials the QCD phase diagram
remains inaccessible to direct lattice simulations because of the sign problem.
Some methods have been developed to tackle the sign problem, such as Taylor expansion \cite{Allton:2002zi}\cite{Gavai:2003mf} and analytic continuation from
imaginary $\mu$ \cite{DElia:2002tig}\cite{deForcrand:2003bz}. These techniques exploit the absence of the sign problem at zero and purely imaginary
chemical potentials. With the former method physical quantities are Taylor expanded around $\mu = 0$,
and the series coefficients are calculated directly on the lattice. With the latter method physical quantities are calculated
at imaginary $\mu$ and then analytically continued to real $\mu$. Here we follow a new method which can be regarded as a combination
of these two approaches. For any given observable we calculate the Taylor series coefficients at zero and purely imaginary chemical potentials.
The Taylor series are merged by using multi-point Pad\'e approximants. Not only these can be analytically continued to real $\mu$,
but information about the analytical structure of the observables can also be obtained by studying the poles of the approximants \cite{Dimopoulos:2021vrk}.

\section{Lee-Yang edge singularities}
The main focus of this work is the study of the critical points of QCD in the complex $\mu_B$ plane. Notably there is a deep connection between 
singularities in the complex plane and phase transitions. Consider for instance the Ising model 
(see \cite{yanglee1952}\cite{fisher1978}\cite{talk_fdr_lat2022}). At $T > T_c$ the free energy has branch cuts on
the imaginary axis for the magnetic field $h$. The branch points are known as Yang-Lee edge singularities. As $T \to T_c$ the branch cuts
pinch the real axis, signaling the presence of a real phase transition. Lee-Yang singularities are particularly relevant for QCD.
From their trajectory one can infer the location of a physical phase transition. Since their presence implies a finite radius of convergence
for Taylor expansions, information can be obtained about the range of validity of the results obtained in the literature by the Taylor
expansion method. Finally from the universal scaling properties of the LYE singularities, information can be obtained about the
non-universal parameters of the theory.

\section{The Roberge-Weiss transition region}
We have investigated by lattice simulations two temperature regimes, one at high temperature ($T=179.5 $ - $ 195.0$ $MeV$) and one
at low temperature ($T=136.1 $ - $ 166.6$ $MeV$). In the high temperature regime we have studied the LYE singularities associated with the RW critical point.
We have performed numerical simulations for $N_f=2+1$ QCD using highly-improved staggered quarks (HISQ) on $36^3 \times 6$ lattices.
We have calculated the baryon number and electric charge cumulants,
$$\chi_{B,Q}^1 \equiv \frac{1}{VT^3} \frac{1}{Z} \frac{\partial Z}{\partial \hat{\mu}_{B,Q}} \mbox{ , }$$
and their first derivative with respect to $\hat{\mu}_B = \frac{\mu_B}{T}$ for $O(10)$ imaginary chemical potentials. 

\begin{figure}[htp]
    \centering
        \includegraphics[width=0.48\textwidth]{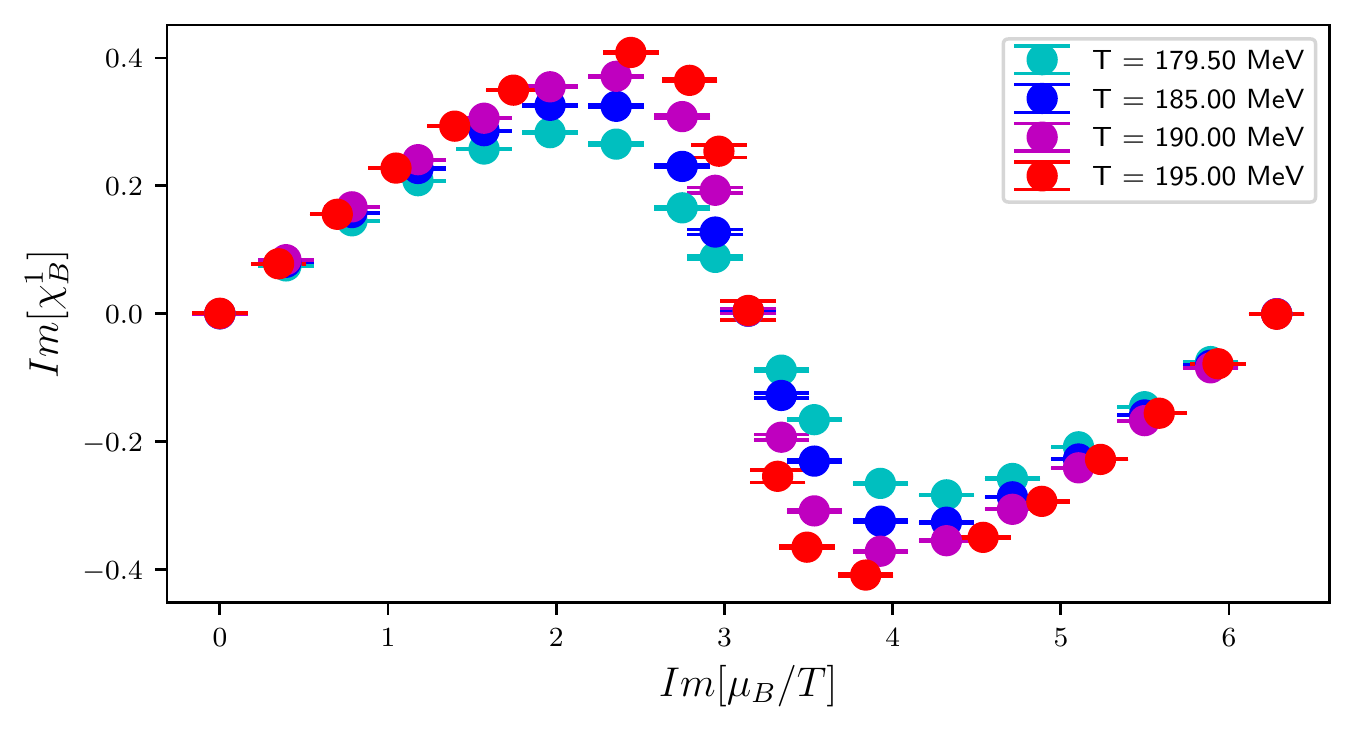} %
        \includegraphics[width=0.48\textwidth]{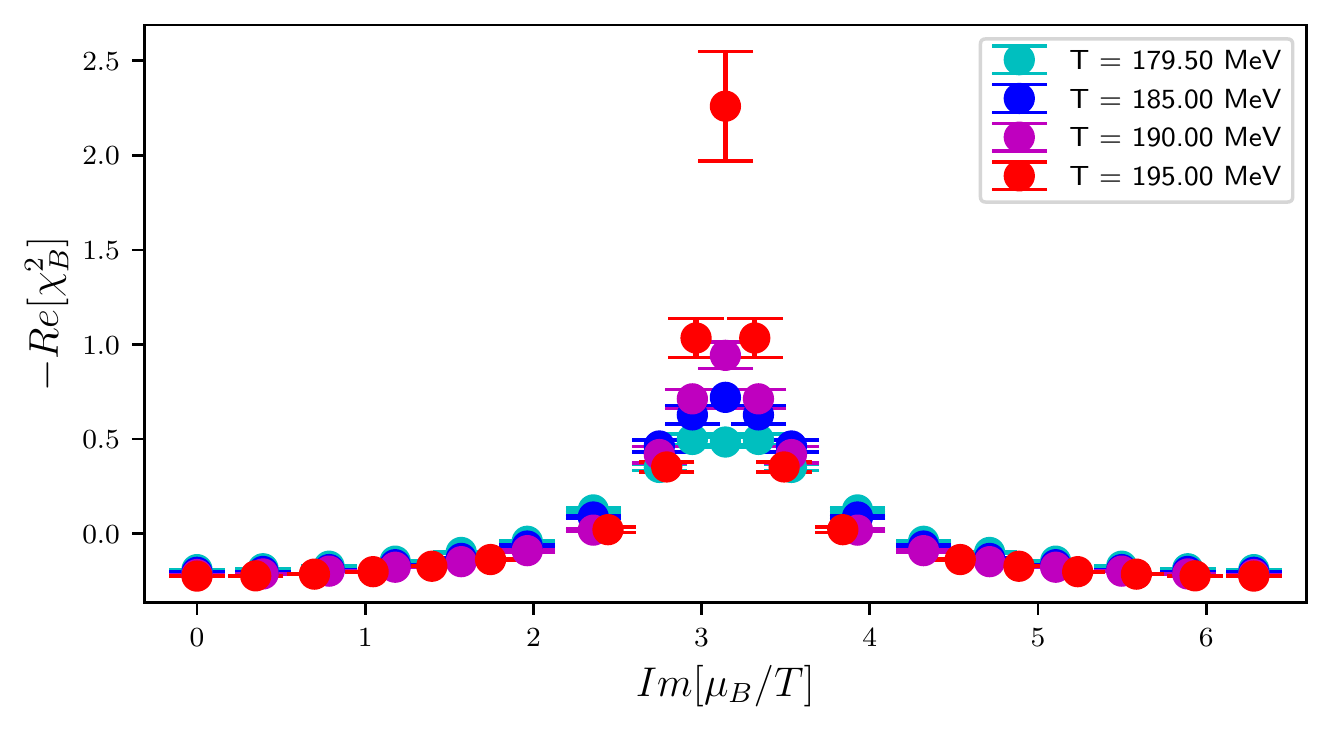} %
    \caption{Imaginary part of $\chi_B^1$ (left) and real part of $\chi_B^2$ (right) as a function of $\hat{\mu}_B$ for different
             temperatures (high temperature regime).}
    \label{fig:rw_obs}
\end{figure}

The calculations were made at four different temperatures $T = 195.0, 190.0, 185.0$ and $179.5$ $MeV$. The highest temperature
is close to the RW temperature that we expect given our choice for the action discretization \cite{Cuteri:2022vwk}.
The numerical results for the baryon number density are shown in fig. \ref{fig:rw_obs}.

\FloatBarrier
\FloatBarrier

\FloatBarrier
\begin{figure}[htp]
    \centering
        \includegraphics[width=0.48\textwidth]{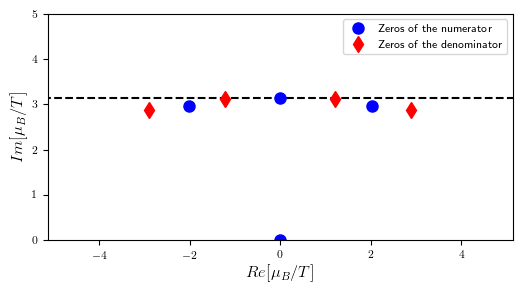} %
        \includegraphics[width=0.48\textwidth]{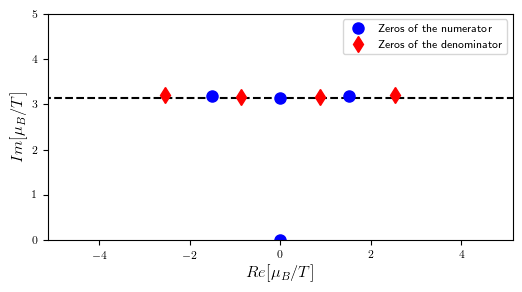} %
        \includegraphics[width=0.48\textwidth]{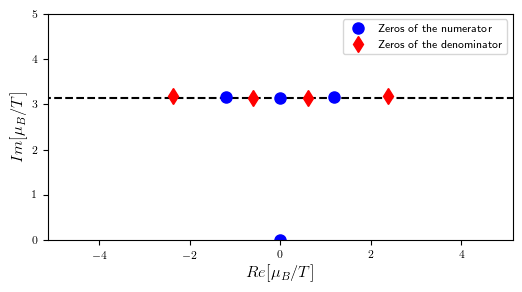} %
        \includegraphics[width=0.48\textwidth]{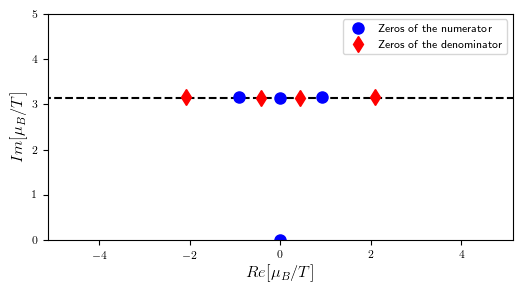} %
    \caption{Singularity structure of the rational approximants at $T = $179.5 $MeV$ (top left), $185.0$ $MeV$ (top right),
            $190.0$ $MeV$ (bottom left) and $195.0$ $MeV$ (bottom right).}
    \label{fig:rw_poles}
\end{figure}
\FloatBarrier

The left picture shows the imaginary part of the baryon number density and the right picture shows the real part of its
derivative with respect to $\hat{\mu}_B$. Different colors correspond to different temperatures and we can see a divergence
emerging at the highest temperature for $\chi_{B}^2$. These data have been approximated by multi-point Pad\'e approximants.
Fig. \ref{fig:rw_poles} shows the singularity structure of the Pad\'e approximants. The zeros of the numerator and denominator
are displayed respectively in blue and red. At $Im(\hat{\mu}_B) = \pi$ we observe an alternation of zeros of the numerator and zeros of
the denominator signaling the presence of branch cuts. The branch points pinch the real axis as the temperature is increased.

\subsection{Scaling analysis}
\begin{figure}[htp]
    \centering
        \includegraphics[width=0.55\textwidth]{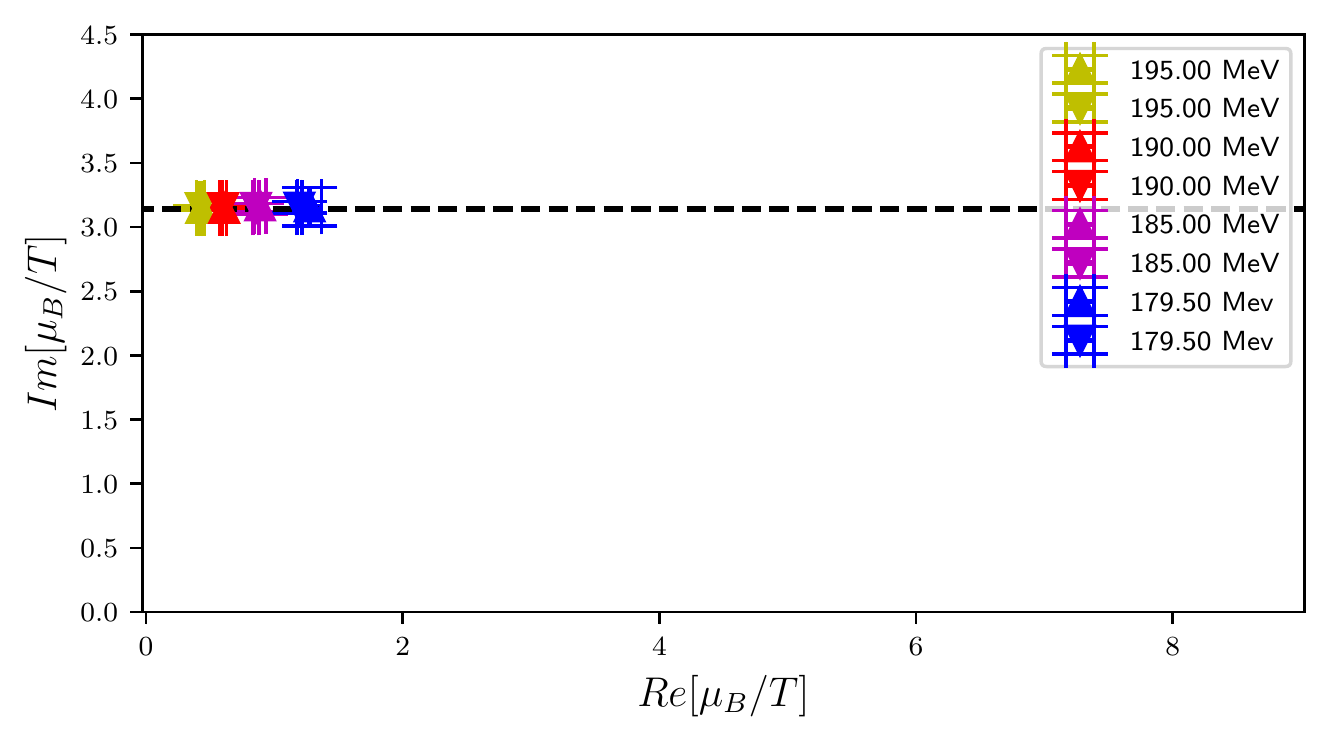} %
        \includegraphics[width=0.5\textwidth]{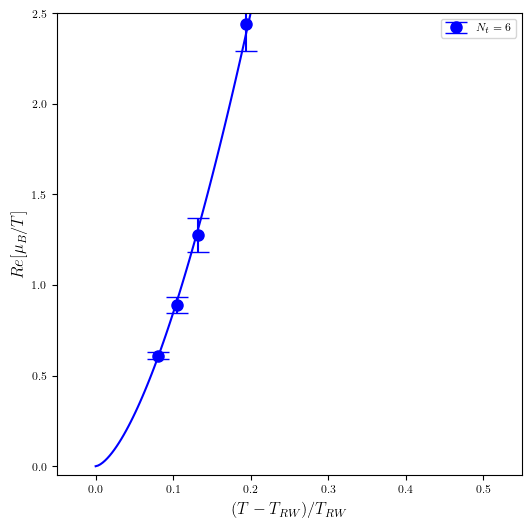} %
    \caption{LYE singularities for different temperatures (top) and scaling fit (bottom).}
    \label{fig:rw_sing}
\end{figure}

In the top picture of fig. \ref{fig:rw_sing} we display for each temperature the closest singularity to the imaginary axis. Different colors
denote different temperatures. Different symbols denote different approximants, respectively the
approximants for $\chi^1_B$ and $\chi^1_Q$. The singularities located by these two observables are in agreement within errors. 
\FloatBarrier
\FloatBarrier

Next we checked whether these singularities could be identified as the LYE singularities associated with the RW endpoint and whether
they follow the critical behaviour expected for the $3d, Z(2)$ universality class. The location of the LYE singularities can be expressed in terms
of the scaling field $z = t/h^{1 / \beta\delta} \equiv |z_c| e^{i\pi / 2\beta\delta}$, where $t$ is the reduced temperature and
$h$ is the symmetry breaking field. This relation makes a connection between the universal parameters
$z_c$, $\beta$, $\delta$ and the non-universal parameters embedded in $t$, $h$ \cite{Connelly:2020gwa} \cite{Johnson:2022cqv} \cite{Dimopoulos:2021vrk}.

For the Roberge-Weiss transition we may introduce the reduced temperature $t = t_0^{-1} \frac{T_{RW} - T}{T_{RW}}$
and the symmetry breaking field $h = h_0^{-1} \frac{\hat{\mu}_B - i \pi}{i \pi}$. One obtains the following scaling law,

$$\hat{\mu}_{LYE} = \pm \pi \left(\frac{z_0}{|z_c|}\right)^{\beta \delta} \left(\frac{T_{RW} - T}{T_{RW}}\right)^{\beta \delta} \pm i \pi \mbox{ , }$$

where $\beta, \delta$ are the universal critical exponents and $z_0 = h_0^{1 / \beta \delta} / t_0$, $T_{RW}$ are non-universal parameters.
In the top picture of fig. \ref{fig:rw_sing} it can be recognized that the imaginary part of the singularities we have found by our Pad\'e analysis is
indeed trivially $i \pi$ within errors. For the real part we can fit the data using the ansatz
  $$\hat{\mu}_{LYE}^R  = a \left(\frac{T_{RW} - T}{T_{RW}}\right)^{\beta \delta} \mbox{ . }$$
By fitting with the $3d, Z(2)$ critical exponents ($\beta \delta = 1.5635$) we obtain  a good fit ($\chi^2 / dof = 0.25$).
From the fit we estimate $T_{RW}(N_\tau = 6) = 206.67(59)$ $MeV$. Using $|z_c| = 2.42(4)$ from ref. \cite{Johnson:2022cqv}
we also find $z_0 = 10.0$-$10.9$.

\subsection{Continuum limit}
\begin{figure}[htp]
    \centering
        \includegraphics[width=0.7\textwidth]{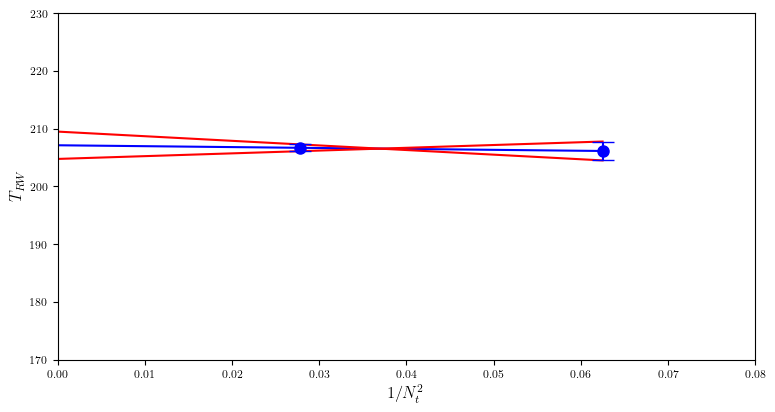} %
    \caption{Estimate for the Roberge-Weiss temperature in the continuum limit.}
    \label{fig:rw_cont}
\end{figure}

We previously conducted a similar analysis using the data obtained from $N_\tau = 4$ lattice simulations \cite{Dimopoulos:2021vrk}.
Using these results and the results from our current analysis we made a crude estimate for the continuum limit.
The result is displayed in fig. \ref{fig:rw_cont}. We obtained $T_{RW}^{cont} = 207.1(2.4)$ $ MeV$.
This is only a preliminary result and a proper continuum extrapolation would require data from $N_\tau=8$ lattice simulations.
Still it is in nice agreement with a previous determination obtained by a different method using a different discretization in ref. \cite{Bonati:2016pwz},
where the authors found $T_{RW}^{cont} = 208(5)$ $ MeV$.

\FloatBarrier
\FloatBarrier
	
\section{The chiral transition and CEP regions}
\begin{figure}[htp]
    \centering
        \includegraphics[width=0.48\textwidth]{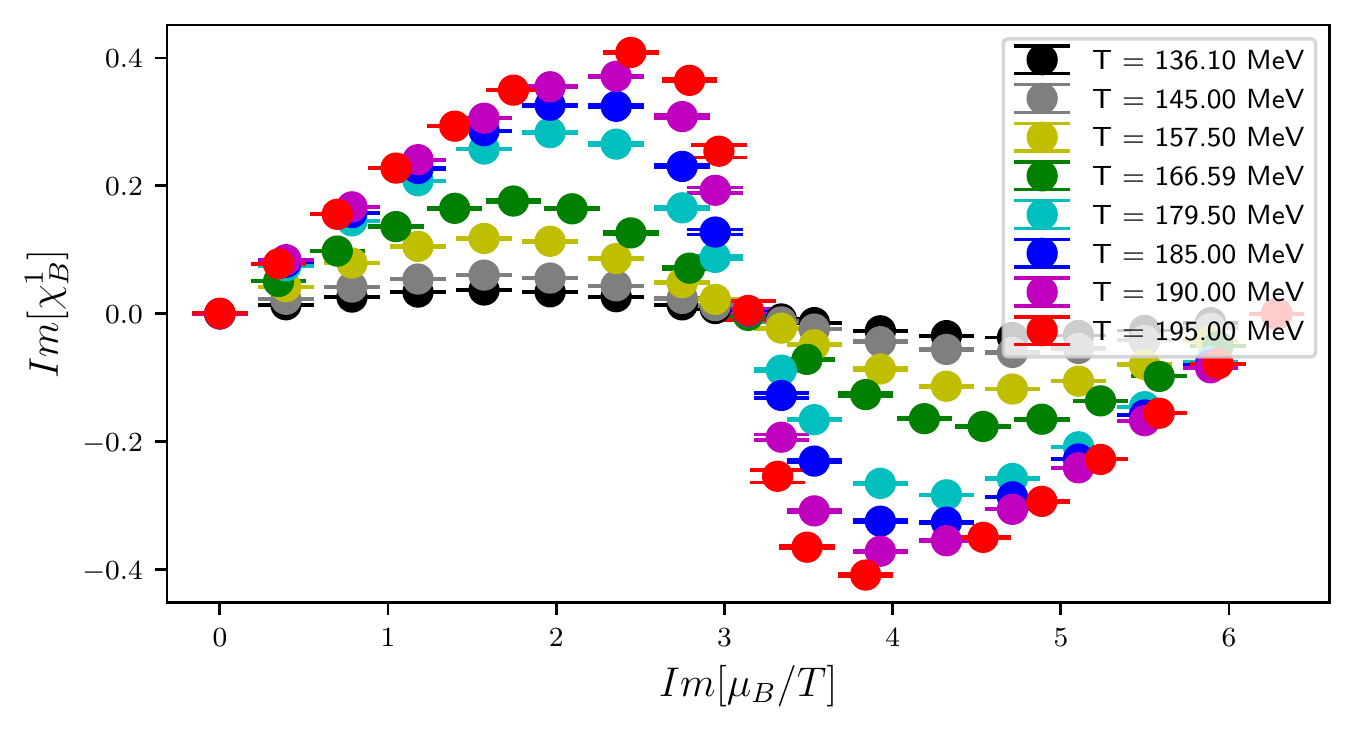} %
        \includegraphics[width=0.48\textwidth]{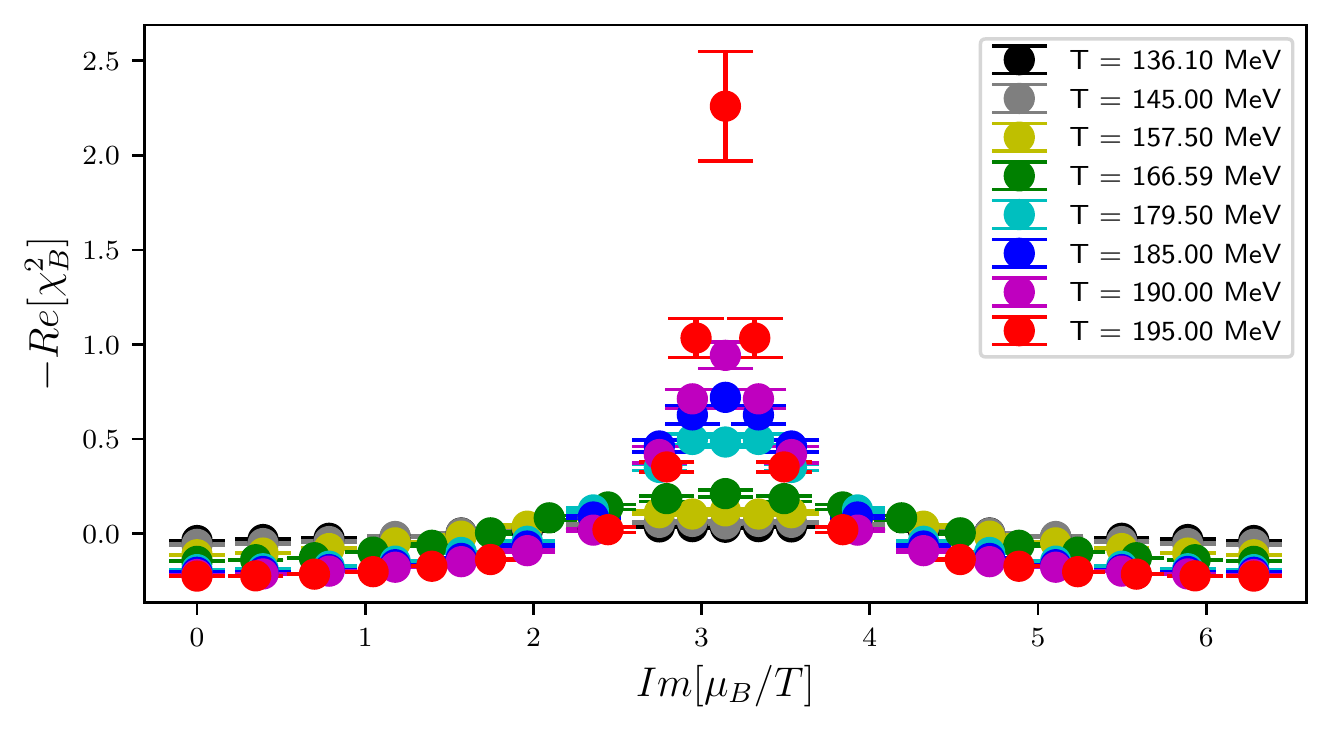} %
   \caption{Imaginary part of $\chi_B^1$ (left) and real part of $\chi_B^2$ (right) as a function of $\hat{\mu}_B$ for different
             temperatures (all temperatures).}
    \label{fig:all_obs}
\end{figure}

In the low temperature regime we have been hunting for the LYE singularities associated with the chiral transition (or possibly the critical endpoint of QCD).
We ran a series of simulations at $T = 166.59, 157.50, 145.00$ and $136.1$ $MeV$.
The numerical results for the baryon number density are shown in fig. \ref{fig:all_obs}.

\FloatBarrier
\FloatBarrier

\begin{figure}[htp]
    \centering
        \includegraphics[width=0.48\textwidth]{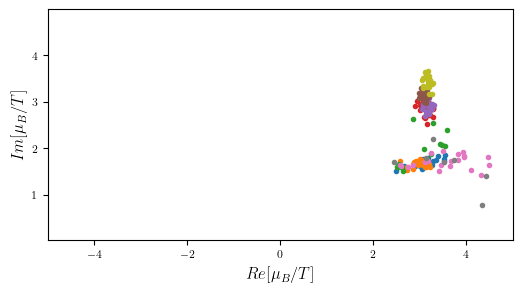} %
        \includegraphics[width=0.48\textwidth]{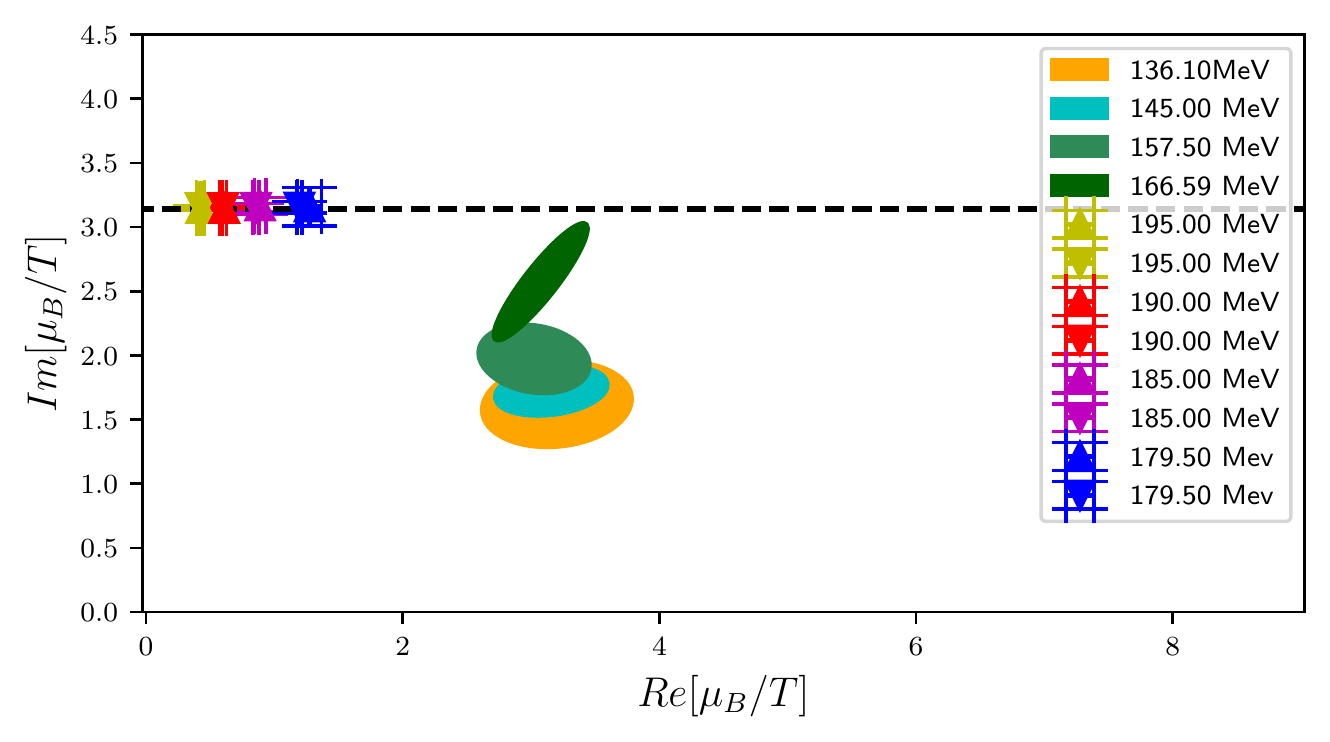} %
    \caption{LYE singularities for different temperatures (right) and scatter plot of the singularities resulting from
             Pad\'e fits over different intervals for $T = 145$ $MeV$ (left).}
    \label{fig:all_sing}
\end{figure}

The low temperature data have been approximated by rational functions, just as we did for the high temperature data.
However in this case we observed a strong interval dependence. This is exemplified in the left picture of fig. \ref{fig:all_sing},
where we plot the singularities resulting from the Pad\'e analysis of the $T=145.00$ $MeV$ data.
Different colors denote fits over different intervals. We can recognize two distinct clusters. The results in the bottom
cluster are from the fits over (small variations of) the $[0, i \pi]$ interval, the results in the
top cluster are from the fits over (small variations of) the full $[0, 2 i \pi]$ interval.
Being located at $Im(\hat{\mu}_B) \approx i \pi$, the top cluster is likely related to the RW endpoint. In the following
we focus on the bottom cluster and we study how it moves as we change the temperature.

\FloatBarrier
\FloatBarrier

\subsection{Scaling analysis}
In the right picture of fig. \ref{fig:all_sing} we summarize the singularities that we have determined in this work.
The uncertainty in the determinations of the singularities in the low temperature regime are given by the $1 \sigma$ confidence ellipses.
These singularities apparently move towards the real axis as the temperature is decreased. Their locations imply
a radius of convergence $\mathcal{R} \approx 3 $ - $ 4$ for Taylor series expansions around $\hat{\mu}_B = 0$ at $T=136 $ - $ 166$ $MeV$.

These singularities might be identified as the LYE singularities associated with the chiral transition or possibly with the critical endpoint of QCD.

\FloatBarrier
\begin{figure}[htp]
    \centering
        \includegraphics[width=1.1\textwidth]{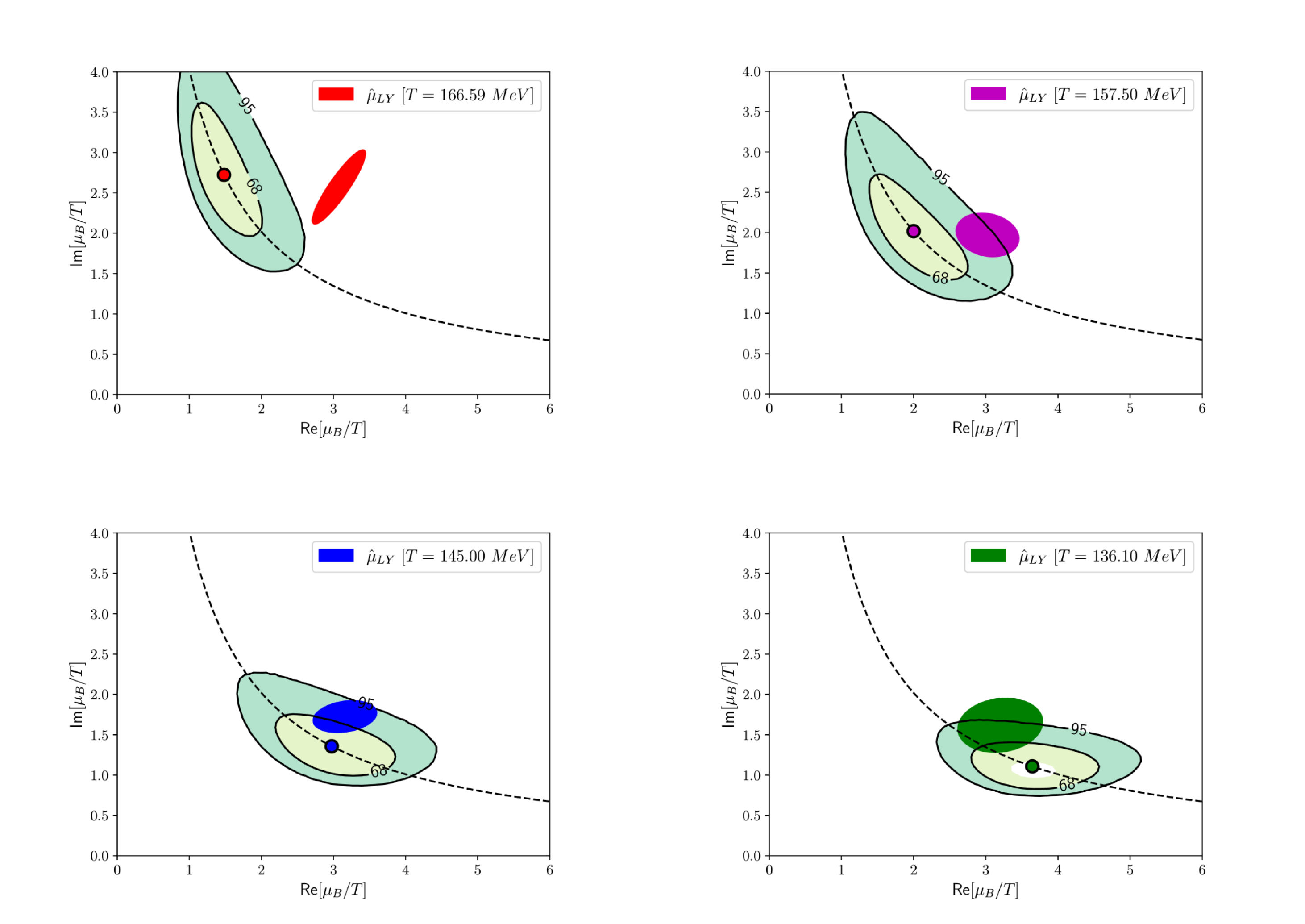} %
    \caption{Comparison with the predictions for the chiral singularities from HotQCD data at temperatures $T = 166.59$ $MeV$ (top left),
             $T = 157.50$ $MeV$ (top right), $T = 145.0$ $MeV$ (bottom left) and $T = 136.10$ $MeV$ (bottom right).}
    \label{fig:chiral_sing_scaling1}
\end{figure}
\FloatBarrier

\FloatBarrier
\begin{figure}[htp]
    \centering
        \includegraphics[width=0.50\textwidth]{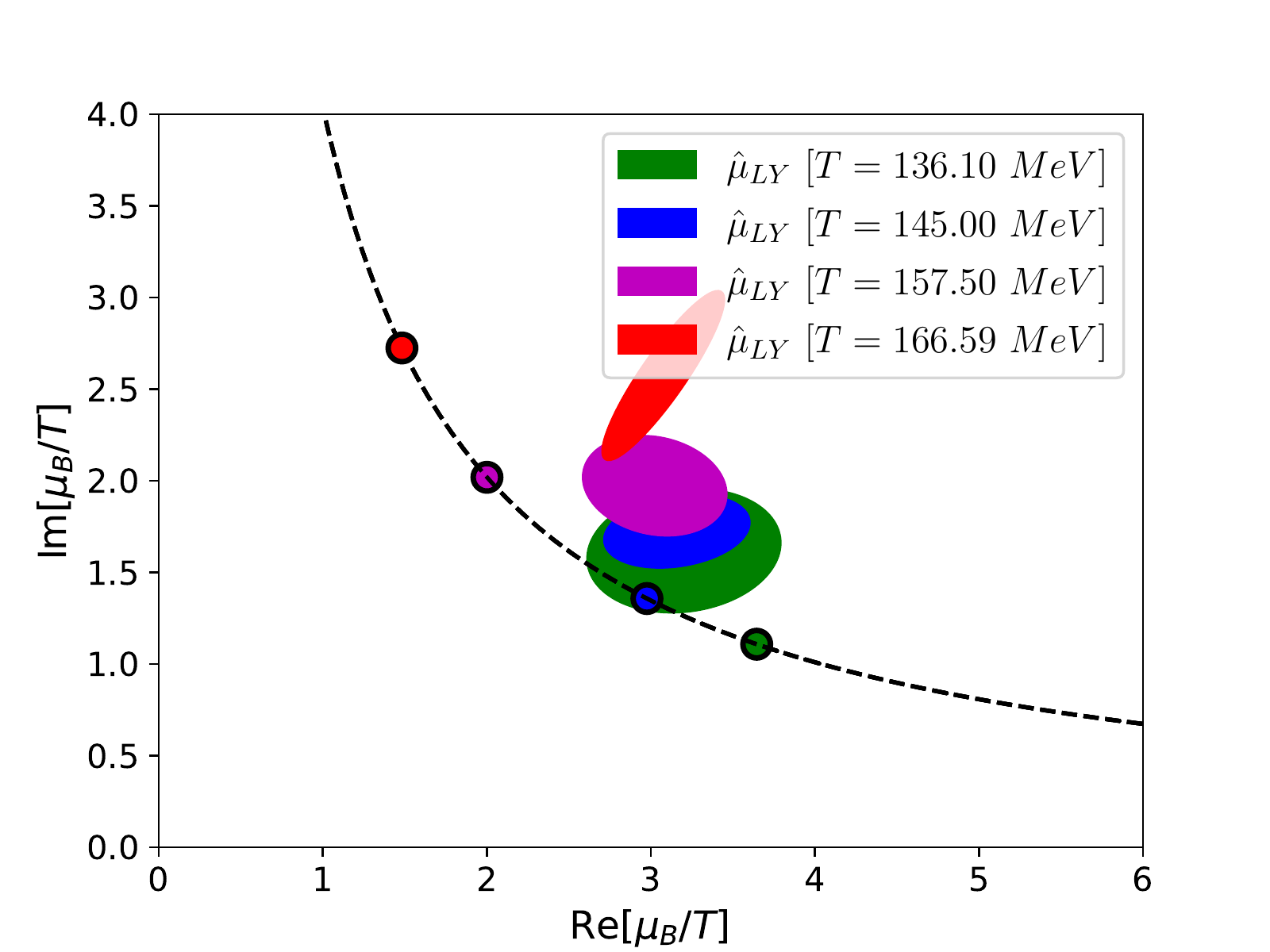} %
        \includegraphics[width=0.45\textwidth]{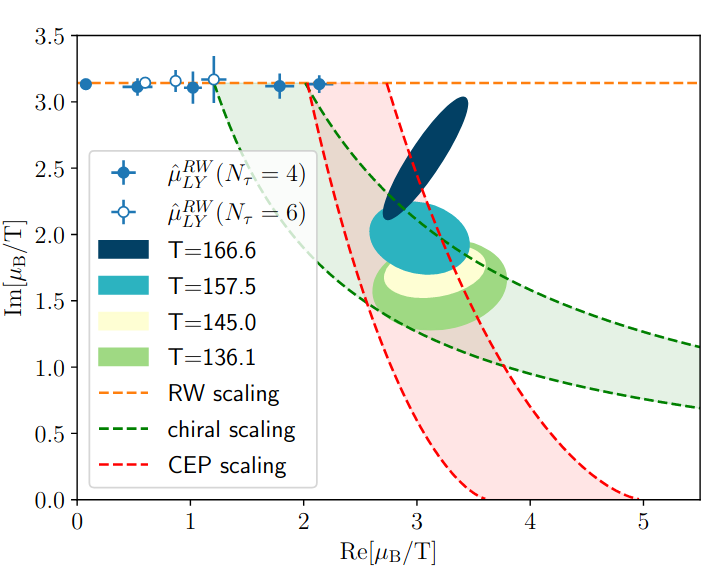} %
    \caption{Overall picture of the comparison with the predictions for the chiral singularities from HotQCD data (left) and summary
             of the expected scaling for the LYE singularities associated with the RW endpoint, with the chiral transition and with
             the CEP (right, figure from ref. \cite{Schmidt:2022ogw}).}
    \label{fig:chiral_sing_scaling2}
\end{figure}
\FloatBarrier

The critical behaviour close to the chiral transition region is expected to follow the behaviour of a theory belonging to the $3d, O(2)$ universality class.
Following ref. \cite{Mukherjee:2019eou} the critical behaviour can be studied by introducing the scaling fields
 $$t = t_0^{-1} \left[\frac{T - T_c}{T_c} + k_2^B \left(\frac{\mu_B}{T}\right)^2\right] \mbox{\phantom{ooo},\phantom{ooo}} h = h_0^{-1} \frac{m_l}{m_s^{phys}} \mbox{ , }$$
where $T_c$ is the critical temperature, $k_2^B$ is the curvature coefficient of the critical line $T_c(\hat{\mu}_B)$ and the symmetry breaking
field $h$ is expressed in terms of the light-to-strange quark mass ratio $\frac{m_l}{m_s^{phys}}$. One can derive the scaling law for the chiral singularities

$$\hat{\mu}_{LYE} = \left[ \frac{1}{k_2^B} \frac{z_c}{z_0} \left(\frac{m_l}{m_s^{phys}}\right)^{\frac{1}{\beta \delta}}  - \frac{T - T_c}{T_c}  \right]^{\frac{1}{2}} \mbox{ . }$$

In fig. \ref{fig:chiral_sing_scaling1} the singularities resulting from the Pad\'e analysis are compared with the predictions obtained by
setting $\frac{m_l}{m_s^{phys}} = 1/27$, by using the non-universal parameters $T_c = 147(6)$ $MeV$, $k_2^B = 0.012(2)$, $z_0=2.35(20)$ from the HotQCD data, 
and by using $|z_c| = 2.032$ from ref. \cite{Connelly:2020gwa}.

The results are in agreement within errors, but if we look at the overall picture of fig. \ref{fig:chiral_sing_scaling2} (left) we observe that
the singularities follow a steeper curve than the one predicted using the best estimates for the non-universal parameters (the dashed line in the
picture).

A possibility remains open that the singularities that we observed are the LYE singularities associated with the critical endpoint of QCD.
In this case the mapping from the non-universal parameters to the universal theory in unknown. 
An approach one may try is to use the linear ansatz \cite{Basar:2021hdf}
 $$t = \alpha_t (T - T_{CEP}) + \beta_t (\mu_B - \mu_{CEP}) \mbox{\phantom{ooo},\phantom{ooo}} h = \alpha_h (T - T_{CEP}) + \beta_h (\mu_B - \mu_{CEP}) \mbox{  }$$
to derive the scaling law
 $$\mu_{LYE} \sim \mu_{CEP} - c_1 (T - T_{CEP}) + i c_2 |z_c|^{-\beta\delta}  (T - T_{CEP})^{\beta \delta} \mbox{ . }$$
Using some reasonable estimates for the non-universal parameters one obtains the qualitative prediction shown as a red band in the right picture
of fig. \ref{fig:chiral_sing_scaling2}. The red band seems to better describe our data than the expected scaling for the chiral singularities
(the green band in the same picture). This conjecture will be explored in future work. A different approach in which LYE singularities
are located by studying the Fourier coefficients of the baryon number density is also being explored \cite{talk_cs_lat2022}.

\subsection{Comparison with $N_\tau=8$ data}
\begin{figure}[htp]
    \centering
        \includegraphics[width=0.65\textwidth]{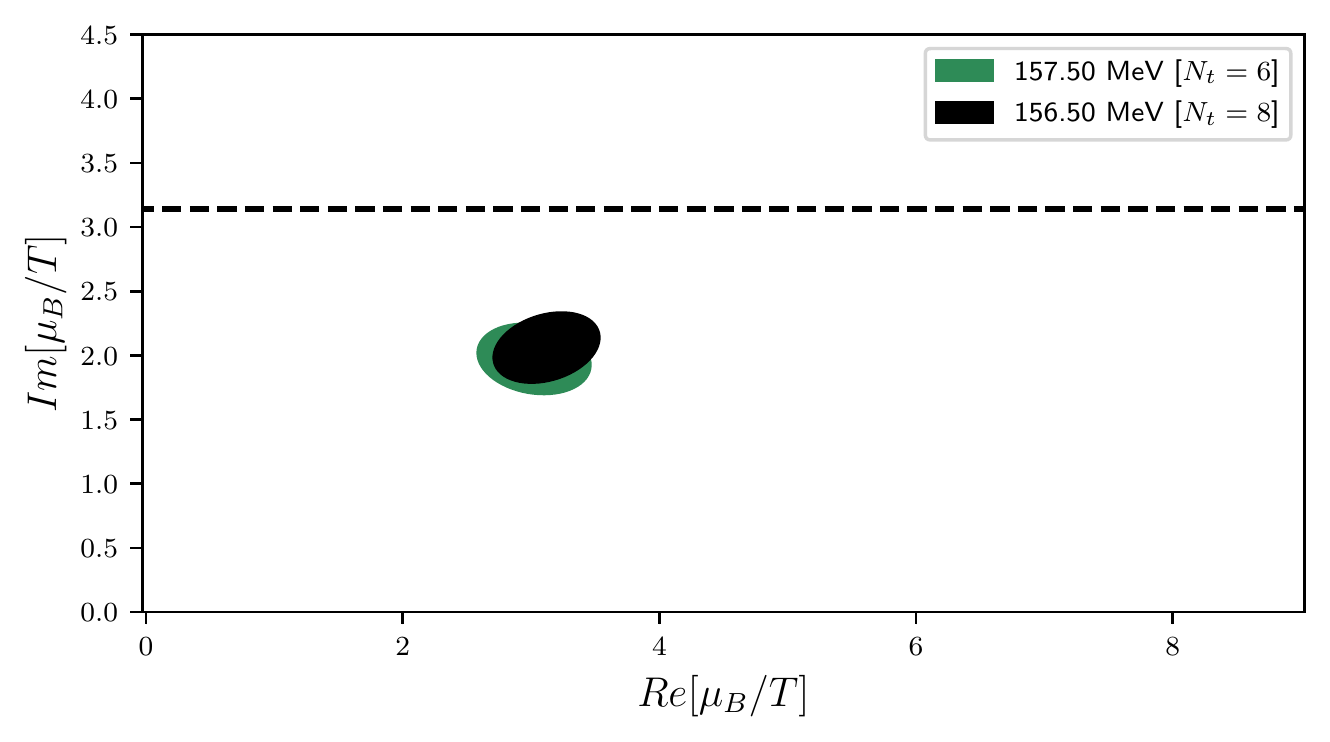} %
    \caption{Comparison between $N_\tau=8$ and $N_\tau=6$ lattices results for the LYE singularities at $T \approx 157$ $MeV$.}
    \label{fig:nt8_sing}
\end{figure}

Finally we ran a separate set of simulations at $T = 156.5$ $MeV$ using $32^3 \times 8$ lattices. The singularity resulting from the Pad\'e analysis
is shown in fig. \ref{fig:nt8_sing}. Also shown is the singularity obtained at a similar temperature ($T = 157.5$ $MeV$) for $N_\tau = 6$ lattices.
The results are in very good agreement despite the different volume. This suggests that both the UV-cutoff effects and the finite-size effects are
negligible within the accuracy of our results.

\FloatBarrier
\FloatBarrier

\section{Conclusions}
We have studied the complex singularities of QCD by a multi-point Pad\'e analysis. In the high temperature regime we 
have identified the LYE singularities associated to the Roberge-Weiss endpoint. These singularities have the 
expected critical behaviour for a transition belonging to the $3d, Z(2)$ universality class. 
By combining these new results and the results
from a previous analysis we have calculated an estimate for the Roberge-Weiss temperature in the continuum limit. 

In the low temperature regime we have found singularities that may be identified as the LYE singularities associated 
with either the chiral transition or the critical endpoint of QCD. Their location imply a radius of convergence $\mathcal{R} = 3 $ - $ 4$ for $T = 136 $ - $ 166$ $MeV$,
which constrains the validity of the Taylor series expansions at $\hat{\mu}_B = 0$.

For future work we plan to extend the study to $N_t = 8$ lattices in order to make a proper continuum extrapolation
for the RW temperature. We also plan to generate data at lower temperatures in order to get a better understanding of
the nature of the complex singularities that we have found at low temperatures.

\section{Acknowledgements}
This work was supported by European Union Horizon 2020 research and innovation
programme under the Marie Sklodowska-Curie grant agreement No 813942 (EuroPLEx) and by
the I.N.F.N. under the research project i.s. QCDLAT. This research used computing resources
made available (i) by CINECA on Marconi and Marconi 100 under both the I.N.F.N-CINECA
agreement and the ISCRA B program, (ii) by the Gauss Centre for Supercomputing on the Juwels GPU nodes at the
Jülich Supercomputing Centre and (iii) by the Bielefeld University on the Bielefeld GPU-Cluster.


\small

\begin{thebibliography}{9}
\bibitem{Guenther:2022wcr}
J.~N.~Guenther,
``An overview of the QCD phase diagram at finite $T$ and $\mu$,''
PoS \textbf{LATTICE2021} (2022), 013
doi:10.22323/1.396.0013
[arXiv:2201.02072 [hep-lat]].

\bibitem{Allton:2002zi}
C.~R.~Allton, S.~Ejiri, S.~J.~Hands, O.~Kaczmarek, F.~Karsch, E.~Laermann, C.~Schmidt and L.~Scorzato,
``The QCD thermal phase transition in the presence of a small chemical potential,''
Phys. Rev. D \textbf{66} (2002), 074507
doi:10.1103/PhysRevD.66.074507
[arXiv:hep-lat/0204010 [hep-lat]].

\bibitem{Gavai:2003mf}
R.~V.~Gavai and S.~Gupta,
``Pressure and nonlinear susceptibilities in QCD at finite chemical potentials,''
Phys. Rev. D \textbf{68} (2003), 034506
doi:10.1103/PhysRevD.68.034506
[arXiv:hep-lat/0303013 [hep-lat]].

\bibitem{DElia:2002tig}
M.~D'Elia and M.~P.~Lombardo,
``Finite density QCD via imaginary chemical potential,''
Phys. Rev. D \textbf{67} (2003), 014505
doi:10.1103/PhysRevD.67.014505
[arXiv:hep-lat/0209146 [hep-lat]].

\bibitem{deForcrand:2003bz}
P.~de Forcrand and O.~Philipsen,
``QCD phase diagram at small densities from simulations with imaginary $\mu$,''
doi:10.1142/9789812704498\_0027
[arXiv:hep-ph/0301209 [hep-ph]].

\bibitem{Dimopoulos:2021vrk}
P.~Dimopoulos, L.~Dini, F.~Di~Renzo, J.~Goswami, G.~Nicotra, C.~Schmidt, S.~Singh, K.~Zambello and F.~Ziesch\'e,
``Contribution to understanding the phase structure of strong interaction matter: Lee-Yang edge singularities from lattice QCD,''
Phys. Rev. D \textbf{105} (2022) no.3, 034513
doi:10.1103/PhysRevD.105.034513
[arXiv:2110.15933 [hep-lat]].

\bibitem{yanglee1952}
T.~D.~Lee and C.~N.~Yang,
``Statistical theory of equations of state and phase transitions. 2. Lattice gas and Ising model,''
Phys. Rev. \textbf{87} (1952), 410-419
doi:10.1103/PhysRev.87.410

\bibitem{fisher1978}
M.~E.~Fisher,
``Yang-Lee Edge Singularity and $\phi^3$ Field Theory,''
Phys. Rev. Lett. \textbf{40} (1978), 1610-1613
doi:10.1103/PhysRevLett.40.1610

\bibitem{talk_fdr_lat2022}
F.~Di~Renzo and S.~Singh,
``Multi-point Pad\'e for the study of phase transitions: from the Ising model to lattice QCD,''
PoS \textbf{LATTICE2022} (2023), 148.

\bibitem{Cuteri:2022vwk}
F.~Cuteri, J.~Goswami, F.~Karsch, A.~Lahiri, M.~Neumann, O.~Philipsen, C.~Schmidt and A.~Sciarra,
``Toward the chiral phase transition in the Roberge-Weiss plane,''
Phys. Rev. D \textbf{106} (2022) no.1, 014510
doi:10.1103/PhysRevD.106.014510
[arXiv:2205.12707 [hep-lat]].

\bibitem{Connelly:2020gwa}
A.~Connelly, G.~Johnson, F.~Rennecke and V.~Skokov,
``Universal Location of the Yang-Lee Edge Singularity in $O(N)$ Theories,''
Phys. Rev. Lett. \textbf{125} (2020) no.19, 191602
doi:10.1103/PhysRevLett.125.191602
[arXiv:2006.12541 [cond-mat.stat-mech]].

\bibitem{Johnson:2022cqv}
G.~Johnson, F.~Rennecke and V.~V.~Skokov,
``Universal location of Yang-Lee edge singularity in classic O(N) universality classes,''
[arXiv:2211.00710 [hep-ph]].

\bibitem{Bonati:2016pwz}
C.~Bonati, M.~D'Elia, M.~Mariti, M.~Mesiti, F.~Negro and F.~Sanfilippo,
``Roberge-Weiss endpoint at the physical point of $N_f = 2+1$ QCD,''
Phys. Rev. D \textbf{93} (2016) no.7, 074504
doi:10.1103/PhysRevD.93.074504
[arXiv:1602.01426 [hep-lat]].

\bibitem{Schmidt:2022ogw}
C.~Schmidt, D.~A.~Clarke, P.~Dimopoulos, J.~Goswami, G.~Nicotra, F.~Di Renzo, S.~Singh and K.~Zambello,
``Detecting critical points from Lee-Yang edge singularities in lattice QCD,''
Acta Phys. Pol. B Proc. Suppl. 16, 1-A52 (2023)
doi:10.5506/APhysPolBSupp.16.1-A52
[arXiv:2209.04345 [hep-lat]].

\bibitem{Mukherjee:2019eou}
S.~Mukherjee and V.~Skokov,
``Universality driven analytic structure of the QCD crossover: radius of convergence in the baryon chemical potential,''
Phys. Rev. D \textbf{103} (2021) no.7, L071501
doi:10.1103/PhysRevD.103.L071501
[arXiv:1909.04639 [hep-ph]].

\bibitem{Basar:2021hdf}
G.~Basar,
``Universality, Lee-Yang Singularities, and Series Expansions,''
Phys. Rev. Lett. \textbf{127} (2021) no.17, 171603
doi:10.1103/PhysRevLett.127.171603
[arXiv:2105.08080 [hep-th]].

\bibitem{talk_cs_lat2022}
C.~Schmidt,
``Fourier coefficients of the net-baryon number density,''
PoS \textbf{LATTICE2022} (2023), 159.

\end{thebibliography}
\end{document}